\documentclass[12pt]{article}

\usepackage{amsfonts,amssymb, amsmath,mathrsfs,hyperref}

\def\Nij{Nijenhuis}

\def\nn{ \nonumber }
\def\bq{ \begin{equation} }
\def\eq{ \end{equation} }
\def\ben{ \begin{eqnarray} }
\def\en{ \end{eqnarray} }

\newtheorem{thr}{Theorem}

\newtheorem{prop}{Proposition}

\newtheorem{exa}{Example}

\newtheorem{defi}{Definition}

\newtheorem{remark}{Remark}

\newenvironment{rem}{\begin{remark} \small \rm}{\end{remark}}

\newenvironment{exam}{\begin{exa} \rm}{\end{exa}}

\begin{document}

%%%%%%%%%%%% TITLE %%%%%%%%%%%%%%

\title{Towards a classification of natural bi-hamiltonian systems.}
\author{ A.V.Tsiganov\\
\\
\it\small St.Petersburg State University, St.Petersburg, Russia\\
\it\small e--mail: tsiganov@mph.phys.spbu.ru}
\date{}
\maketitle

\begin{abstract}
For construction and classification of the natural integrable
systems we propose to use a criterion of separability in
Darboux--\Nij\ coordinates, which can be tested without an a priori
explicit knowledge of these coordinates.
\end{abstract}

\section{Introduction}
\setcounter{equation}{0}

The search of integrable dynamical systems is one of the most
fascinating branch of classical physics. Integrable systems are
quite rare and still only a few examples are known. In this paper we
will present a new direct method for construction of natural
integrable systems on  $\omega N$ bi-hamiltonian manifolds.

A bi-hamiltonian manifold is a smooth  manifold $M$ endowed with a
pair of compatible Poisson tensors $P$ and $P^{\,\prime}$ associated
with the Poisson brackets  $\{.,.\}$ and $\{.,.\}'$ respectively.
The class of manifolds we will consider are particular
bi-Hamiltonian manifolds where one of the two Poisson tensors in non
degenerate and thus defines a symplectic form $\omega=P^{-1} $ and a
recursion operator $N=P^{\,\prime}P^{-1}$.

Dynamical integrable  systems on $M$ with functionally independent
integrals of motion $H_1,\ldots,H_n$ in the  bi-involution
\bq\label{bi-ham}
\{H_i,H_j\}=0,\quad\mbox{and}\quad \{H_i,H_j\}'=0,\qquad\mbox{for
all}\, i,j
\eq
will be called bi-hamiltonian systems for brevity.

A Hamiltonian system is  natural if its Hamiltonian is a sum of a
positive-definite kinetic energy and a potential.  For instance, on
the symplectic manifold $M\simeq \mathbb R^{2n}$, i.e. in the
Euclidean space, the natural Hamiltonian is equal to
\bq\label{nat-ham}
H=\sum_{i=1}^n p_i^2+V(q)\,.
\eq
In the known direct method to find natural integrable systems
\cite{hiet87} we substitute natural Hamiltonian and some ansatz for
other integrals of motion into the equations (\ref{bi-ham}) and try
to solve these equations.

According to \cite{fmp01,fp02} integrals of motion $H_1,\ldots,H_n$
are in bi-involution (\ref{bi-ham}) if  they are separable in
Darboux-\Nij\ coordinates which are canonical with respect to
$\omega$ and diagonalize $N$. So, instead of classification of
bi-hamiltonian systems we can classify Hamiltonians whose associated
Hamilton-Jacobi equations can be solved by separation of variables
in Darboux-\Nij\ coordinates.

 A test for separability of a given Hamiltonian $H$ in a given coordinate system
was proposed by  Livi-Civita in 1904 \cite{lc04}. Using
Darboux--\Nij\ coordinates associated with $N$ the Livi-Civita
criteria  may be rewritten in the following form \cite{bfp03}
\bq\label{sep-cond}
d(N^*dH)(\mathcal D_H,\mathcal D_H)=0\,.
\eq
Here $\mathcal D_H$ is distribution generated iteratively by the
action of $N$ on the vector field $X_H$.

The relevance of a new  form of the Livi-Civita criteria is that
condition (\ref{sep-cond})  can be checked in any coordinate system.
So, for a given Hamiltonian $H$ we can look at (\ref{sep-cond}) as
one of the equations to determine torsion free tensor $N$ and hence
the separated variables. This scheme, in its basic features, has
already been considered in the literature and applied to various
systems, see \cite{bfp03,fmp01,fp02,imm00,mag97,mt97,mt02,ts05a} and
references within.

In this paper we will take a different logical standpoint. We
propose to solve equation (\ref{sep-cond}) with respect to a
torsionless tensor $N$ and a natural Hamiltonian $H$ simultaneously.
We will prove that it is indeed possible and, actually, easy to
solve such an equation by means of a couple of natural Ans\"{a}tze.
As an example,  we reproduce a lot of known natural integrable
systems on the plane with cubic and quartic integrals of motion
\cite{hiet87}. The corresponding Poisson pencils and tensors $N$ are
new.

%%%%%%%%%%%%%%%%%%%%%%%%%%%%%%%%%%%%%%%%%%%%%%%%%%%%%%%%%%%%%%%%%%%%%%%%%%%
\section{Bihamiltonian geometry and separation of variables}
 \setcounter{equation}{0}

In this section we sketch the main points of the  separation of
variables theory for bi-hamiltonian manifolds, referring to
\cite{bfp03,fmp01,fp02,imm00} for complete proofs and more detailed
discussions.

\subsection{ $\omega N$ manifolds}

The phase spaces of integrable systems will be identified with
$\omega N$-manifolds \cite{fmp01,fp02} or Poisson-Nijenhuis
manifolds \cite{ksm90,mag97} endowed with a symplectic form $\omega$
and a torsion free tensor $N$ satisfying certain compatibility
conditions.
\begin{defi}
An $\omega N$ manifold  is a bi-hamiltonian manifold in which one of
the Poisson tensor (say, $P$)  is non degenerate, i.e. $\det P\neq
0$\,.
\end{defi}
 Therefore,
$M$ is endowed with a symplectic form $\omega=P^{-1}$ and with a
tensor field
\bq\label{recc-op} N=P^{\,\prime}P^{-1},
\eq
which is recursion operator (formally). The operator $N$ is called a
\Nij\ tensor \cite{mag97} or a hereditary operator \cite{ff81} as
well.

In order to construct operator $N$ by the rule (\ref{recc-op}) we
have to get a pair of compatible Poisson tensors $P$ and
$P^{\,\prime}$.

\begin{defi}The Poisson tensors $P$ and $P^{\,\prime}$ are compatible  if every
linear combination of them is still a Poisson tensor. The
corresponding linear combination $P^{\lambda}= P+\lambda
P^{\,\prime}$, $\lambda\in \mathbb R$, is a Poisson pencil.
\end{defi}

Recall a bivector $P^{\,\prime}$ on $M$ is said to be a Poisson
tensor if the Poisson bracket associated with it
\[
\{f(z),g(z)\}^\prime=\sum_{i,k=1}^{2n}
P_{ik}^\prime(z)\dfrac{\partial f(z)}{\partial z_i}\dfrac{\partial
g(z)}{\partial z_k}
\]
satisfies the Jacobi identity, that is, if for  all  $f,g$ and $h\in
C^{\infty}( M)$
\bq\label{Jac-id2}
\{f,\{g,h\}^\prime\}^\prime+\{g,\{h,f\}^\prime\}^\prime+\{h,\{f,g\}^\prime\}^\prime=0\,.
\eq
Here $z=(z_1,\ldots,z_{2n})$ is a point of $M$ in some coordinate
system.

From the Jacobi identity we can get a system of equations on the
entries $P_{ik}^{\,\prime}(z)$ of the tensor $P^{\,\prime}$. So, in
order to construct the Poisson tensor $P^{\,\prime}$ we have to
solve algebraic equations
\[P^{\,\prime}_{ij}=-P^{\,\prime}_{ji},\qquad
i,j=1,\ldots,2n,\] and  partial differential equations
\bq\label{jac1}
\sum_{m=1}^{2n} \left(P^{\,\prime}_{im}\dfrac{\partial
P^{\,\prime}_{jk}}{\partial z_m}+
                   P^{\,\prime}_{jm}\dfrac{\partial P^{\,\prime}_{ki}}{\partial z_m}+
                   P^{\,\prime}_{km}\dfrac{\partial P^{\,\prime}_{ij}}{\partial z_m}
\right)=0,\qquad i,j,k=1,\ldots,2n,
\eq
with respect to some unknown functions $P_{ik}^\prime(z)$.

In order to construct the Poisson pencil on $\omega N$ manifold
associated with a given tensor $P$ we have to solve equations
(\ref{jac1}) simultaneously with the equations
\bq\label{jac2}
\sum_{m=1}^{2n} \left(P^{\lambda}_{im}\dfrac{\partial
P^{\lambda}_{jk}}{\partial z_m}+
                   P^{\lambda}_{jm}\dfrac{\partial P^{\lambda}_{ki}}{\partial z_m}+
                   P^{\lambda}_{km}\dfrac{\partial P^{\lambda}_{ij}}{\partial z_m}
\right)=0,\qquad i,j,k=1,\ldots,2n,
\eq
which follows from the Jacobi identity for the Poisson bracket $\{.
,. \}_\lambda=\{. ,. \}+\lambda\{. ,. \}^\prime$, where $\lambda$ is
an arbitrary numerical parameter.

For any solution $P^{\,\prime}$ of the equations
(\ref{jac1}-\ref{jac2}) there is free torsion tensor $N$
(\ref{recc-op}).
\begin{thr} \cite{mag97}
The \Nij\  torsion of $N$,
\bq\label{kr-n}
T_N(X,Y)=[NX,NY]-N\Bigl( [NX,Y]+[X,NY]-N[X,Y] \Bigr)=0,
\eq
vanishes as a consequence of the compatibility between $P$ and
$P^{\,\prime}$.
\end{thr}
In a given coordinate system on $M$ entries of the  \Nij\ torsion of
$N$ are equal to
\[
T^{i}_{jk}(N)=\sum_{m=1}^{2n}\left( \dfrac{\partial N^i_k}{\partial
z_m}N^m_j-\dfrac{\partial N^i_j}{\partial z_m}N^m_k +\dfrac{\partial
N^m_j}{\partial z_k}N^i_m-\dfrac{\partial N^m_k}{\partial z_j}N^i_m
\right)\,.
\]
In general there are infinitely many solutions of the system of
equations (\ref{jac1}-\ref{jac2}) or  (\ref{kr-n}).

In order to identify $2n$ separated variables with Darboux-\Nij\
coordinates we have to consider a special class of $\omega N$
manifolds.
\begin{defi}
A $2n$-dimensional $\omega N$ manifold $M$ is said to be semisimple
if its recursion operator $N$ has, at every point, $n$ distinct
eigenvalues $\lambda_1,\ldots, \lambda_n$. It is called regular if
eigenvalues of $N$ are functionally independent on $M$.
\end{defi}
Of course, equations (\ref{jac1}-\ref{jac2}) also have infinitely
many solutions on semisimple regular $\omega N$ manifolds. In order
to get a bounded set of the solutions we have to fix  form of the
admissible Poisson tensors $P^{\,\prime}$.

\begin{exam}
Let us consider canonical Poisson tensor in the  Euclidean space
\bq\label{can-Poi}
 P=\left(%
\begin{array}{cc}
 0 & {\mathrm I} \\
 -{\mathrm I} & 0 \\
\end{array}%
\right),\qquad {\mathrm I}=\mbox{\rm diag}(1,\ldots,1)
\eq
associated with the usual Poisson bracket for arbitrary functions
$f,g\in C^{\infty}(M)$
\bq \label{poi-Rn}
\{f,g\}=\sum_{i=1}^n\left(\frac{\partial f}{\partial
p_i}\frac{\partial g}{\partial q_i}-\frac{\partial f}{\partial
q_i}\frac{\partial g}{\partial p_i}\right)\,.
\eq
At $n=2$ there are two nontrivial solutions of the system of
equations (\ref{jac1}-\ref{jac2})
\bq\label{kepl-tenz}
P^{\,\prime}=\left(%
\begin{matrix}
0 & 0 & 0 & q_1 \\
 * & 0 & q_1& 2q_2 \\
 * & * & 0 & -p_1 \\
 * & * & * & 0
\end{matrix}%
\right)\quad\mbox{\rm and}\quad
P^{\,\prime}=\left(%
\begin{matrix}
 0 & 0 & a_1-q_1^2 & -q_1q_2 \\
 * & 0 & -q_1q_2& a_2-q_2^2 \\
 * & * & 0 & -q_1p_2+q_2p_1 \\
 * & * & * & 0
\end{matrix}%
\right)\,,
\eq
which are matrix polynomials in variables $z=(p,q)$. The \Nij\
torsion of the corresponding recursion operator $N$ (\ref{recc-op})
vanishes and its eigenvalues are distinct in the both cases.
\end{exam}

\subsection{The Darboux--\Nij\  coordinates}

 According to \cite{br93} the recursion operator $N$ exists for overwhelming majority
of integrable by Liouville systems. Therefore  classification of the
Poisson pencils has to coincide with the classification of
integrable systems, which has a more rich history \cite{hiet87}.
This connection is based on the separation of variables method in
which separated variables for the Hamilton-Jacobi equation are
identified with  Darboux--\Nij\ coordinates \cite{mag90}.
\begin{defi}
A set of local coordinates $(x_i, y_i)$ on an $\omega N$ manifold is
a set of  Darboux--\Nij\  coordinates if they are canonical with
respect to  the symplectic form
\[
\omega=P^{-1}=\sum_{i=1}^n dy_i\wedge dx_i\ ,
\]
and put the recursion operator $N$ in diagonal form
\bq\label{xy-pedr}
 N=\sum_{i=1}^n\lambda_i\left(
\dfrac{\partial }{\partial  x_i}\otimes dx_i+ \dfrac{\partial
}{\partial y_i}\otimes dy_i\right),
\eq
This means that the only nonzero Poisson brackets are $
\{x_i,y_j\}=\delta_{ij}$ and $\{x_i,y_j\}'=\lambda_i\delta_{ij}$.
\end{defi}
According to (\ref{xy-pedr}) differentials of the  Darboux--\Nij\
coordinates  span an eigenspace of the adjoint operator $N^*$
\bq\label{qv-pedr}
N^*dx_i=\lambda_idx_i,\qquad N^*dy_i=\lambda_idy_i\,.
\eq
As a consequence of the vanishing of the \Nij\ torsion of $N$, the
eigenvalues $\lambda_i$ always satisfy
\[
N^* d\lambda_i=\lambda_i d\lambda_i\,.
\]
It allows us to find very easy one special family of Darboux--\Nij\
 coordinates.
\begin{thr} \cite{fmp01,fp02}
In a neighborhood of any point of a regular $\omega N$ manifold
where the eigenvalues of $N$ are distinct it is possible to find by
quadratures $n$ functions $\mu_1, \ldots, \mu_n$ that, along with
the eigenvalues $\lambda_1,\ldots,\lambda_n$, are  Darboux--\Nij\
coordinates.
\end{thr}
In this case the coordinates $\lambda_j$ can be computed
algebraically as the roots of the minimal polynomial of $N$
\[
\Delta(\lambda)=\Bigl(\det(N-\lambda\, {\mathrm
I})\Bigr)^{1/2}=\prod_{j=1}^n (\lambda-\lambda_j)\,.
\]
On the contrary, the conjugated momenta $\mu_j$ must be computed (in
general) by a method involving quadratures \cite{fmp01,fp02}.

\begin{exam} For the first Poisson tensor in
(\ref{kepl-tenz}) the minimal characteristic polynomial of $N$ is
equal to
\[
\Delta(\lambda)=\lambda^2-q_2\lambda-\dfrac{q_1}{4}=(\lambda-\lambda_1)(\lambda-\lambda_2)\,.
\]
The corresponding special Darboux--\Nij\ coordinates
\[
\lambda_{1,2}=\dfrac12\left(q_2\pm\sqrt{q_1^2+q_2^2}\right)\,,\qquad
\mu_{1,2}=p_2-\dfrac{p_1}{q_1}\left(q_2\pm\sqrt{q_1^2+q_2^2}\right)
\]
coincide with the  parabolic coordinates on the plane and
 their conjugated momenta.

For the second Poisson tensor in (\ref{kepl-tenz}) the corresponding
minimal polynomial
\[
\Delta(\lambda)=\lambda^2+(q_1^2+q_2^2-a_1-a_2)\lambda+a_1a_2-a_2q_1^2-a_1q_2^2
=(\lambda-\lambda_1)(\lambda-\lambda_2)
\]
is a generating function of the  elliptic coordinates
$\lambda_{1,2}$ on the plane \cite{ben93}.
\end{exam}

\subsection{Integrable systems in the Jacobi method}

 The  Darboux--\Nij\
coordinates $(x_i,y_i)$ or $(\lambda_i,\mu_i)$ may be identified
with the separated variables for a huge family of integrable
systems. In order to construct these integrable systems in framework
of the Jacobi method we can use the following theorem.
\begin{thr}
Let $(\lambda_i,\mu_i)$ are canonical coordinates
$\{\lambda_i,\mu_j\}=\delta_{ij}$. The product of $n$
one-dimensional Lagrangian submanifolds
\bq\label{sep-eq}
\mathcal
C_i:\quad\Phi_i(\lambda_i,\mu_i,\alpha_1,\ldots,\alpha_n)=0\qquad\mbox{
with}\qquad \det\left\|\dfrac{\partial
\Phi_i(\lambda_i,\mu_i)}{\alpha_j}\right\|\neq 0\,
\eq
is an $n$-dimensional Lagrangian submanifold $\mathcal F(\alpha) =
\mathcal C_1\times\cdots\times \mathcal C_n$. The solutions of the
separated equations (\ref{sep-eq}) with respect to $n$ parameters
$\alpha_k$ are functionally independent integrals of motion $H_k=
\alpha_k(\lambda,\mu)$ in  involution.
\end{thr}
The main problem in the Jacobi approach is to choose  separated
equations $\Phi_i=0$  for which the Hamilton function $H=H_1$ has a
natural form (\ref{nat-ham}) in the initial physical variables
$(p,q)$.

We can reformulate this problem by using  the separability condition
(\ref{sep-cond}) which implies that the distribution $\mathcal D_H$
is integrable. So, there exist $n$ independent local functions
$H_1,\ldots,H_n$ that are constant on the leaves of $\mathcal D_H$.
\begin{thr} \cite{fmp01,fp02}
Let $M$ be a semisimple regular $\omega N$ manifold. The functions
$H_1,\ldots,H_n$ on $M$  are separable in Darboux--\Nij\ coordinates
if and only if they are in bi-involution (\ref{bi-ham}). The
distribution $\mathcal D$ tangent to the foliation defined by
$H_1,\ldots, H_n$ is Lagrangian with respect to $\omega$ and
invariant with respect to $N$.
\end{thr}
The invariance of the Lagrangian distribution $\mathcal D$  means
that  there exists a  matrix $F$ with eigenvalues
$(\lambda_1,\dots,\lambda_n)$ such that
\bq\label{F-pedr}
N^*dH_j=\sum_{k=1}^n  F_{jk}dH_k,\qquad j=1,\ldots,n.
\eq
Here $F$ is the control matrix with respect to integrals of motion
$H_1,\ldots,H_n$, which form  the \Nij\ chain \cite{fmp01,fp02}.

Using this theorem we can formulate a new method to find natural
integrable systems. Namely, substituting a torsionless tensor $N$,
some control matrix $F$ and a natural Hamiltonian $H=H_1$
(\ref{nat-ham}) into the $n$ relations (\ref{F-pedr}),  one gets a
system of $n$ differential equations on the potential $V(q)$ and on
the remaining integrals of motion $H_2,\ldots,H_n$. Solving these
equations one gets a natural integrable system separable in
Darboux--\Nij\ coordinates. In order to avoid a lot of intermediate
calculations we can solve equations
 (\ref{F-pedr}) together with equations (\ref{jac1},\ref{jac2}).

The proposed construction  inherits the main disadvantage of the
Jacobi method because we have to choose control matrix $F$ for which
equations (\ref{F-pedr}) have non-trivial solutions.

\begin{exam} If $H_k=\frac{1}{2k}\mbox{\rm tr} N^k$ then
the control matrix in (\ref{F-pedr}) is equal to
\bq\label{f-num}
F=\left(%
\begin{matrix}
  0 & 1 & 0 &\cdots & 0 \\
  0 & 0 & 1 & \cdots &0 \\
  \vdots &  & \cdots &0 & 1 \\
  \sigma_n & \sigma_{n+1} & \cdots & &\sigma_1
\end{matrix}%
\right).
\eq
Here $\sigma_k$ are the elementary symmetric polynomials of degree
$k$ on the eigenvalues of $N$.

In this case relations (\ref{F-pedr}) are the Lenard-Magri
recurrence relations
\bq\label{bi-chain} N^*\,d{H}_i=d{H}_{i+1}\,\qquad\mbox{\rm or}
\qquad P^{\,\prime}d{H}_i=Pd{H}_{i+1},
\eq
such that the corresponding vector fields $X_{H_i}$ are
bi-hamiltonian \cite{imm00,mag97}.
\end{exam}

\begin{rem}
It is known that for finite-dimensional systems the bi-hamiltonian
property of the vector fields is a very strong condition and too
restrictive for  construction of natural integrable systems
\cite{fp02,mt97}. Of course,  we could easy solve equations
(\ref{F-pedr}) with such matrix $F$ (\ref{f-num}), but for the
majority of the known Poisson pencils system of equations
(\ref{F-pedr}) or (\ref{bi-chain}) is an inconsistent system if
$H_1=H$ is a natural Hamiltonian (\ref{nat-ham}).
\end{rem}

\subsection{The uniform St\"ackel systems}
The equations (\ref{F-pedr}) do not give explicit information on the
form of the separated equations (\ref{sep-eq}). In order to restrict
a set of the possible separated equations $\Phi_i=0$ we will
consider the following family of the separated equations
\bq\label{st-eq}
\Phi_i(\mu_i,\lambda_i,\alpha_1,\ldots,\alpha_n) =\sum_{j=1}^n
S_{ij}(\lambda_i)\,\alpha_j-s_i(\mu_i,\lambda_i)=0\,,\qquad
H_i=\alpha_i\,,
\eq
where $S$ is a non-degenerate St\"ackel matrix and $s$ is  a
St\"ackel vector.  The entries $S_{ij}$ and $s_i$ depend on a pair
of the  separated variables $\lambda_i$ and $\mu_i$ only.

\begin{thr} \cite{fmp01,fp02}
Let $H_1,\dots,H_n$ are independent integrals of motion in
bi-involution (\ref{bi-ham}) defining a bi-Lagrangian foliation on a
regular semisimple $\omega N$ manifold. If the control matrix $F$
satisfies
\bq\label{42-pedr}
N^* dF= Fd F,\qquad\mbox{\it that is}\qquad N^*d F_{ij}=\sum_{k=1}^n
F_{ik}d F_{kj},\qquad i,j=1,\ldots,n,
\eq
then the functions $H_1,\dots,H_n$ are St\"ackel separable in the
Darboux--\Nij\ coordinates.
\end{thr}
In this case the St\"ackel matrix $S$ (\ref{st-eq}) is defined by
\[ F=S^{-1}\Lambda S,\qquad \Lambda=\mbox{\rm
diag}(\lambda_1,\ldots,\lambda_n).
\]
Construction of the corresponding St\"ackel vector $s$ is not so
algorithmic.

This theorem could be  very useful for a construction of natural
integrable systems of the St\"ackel type. Namely, substituting  a
(1,1) tensor  $N$ and a natural Hamiltonian $H=H_1$ (\ref{nat-ham})
into the relations (\ref{kr-n}), (\ref{F-pedr}) and (\ref{42-pedr})
one gets the closed system of algebro-differential equations
\bq\label{gen-sys}
T_N=0,\qquad N^*dH=FdH,\qquad N^* dF= Fd F,
\eq
on potential $V(q)$, integrals of motion $H_2(p,q),\ldots,H_n(p,q)$,
entries $N_{ij}(p,q)$ of the recursion operator and entries
$F_{ij}(p,q)$ of the control matrix.

Unfortunately,  we can not solve equations (\ref{gen-sys}) in
general. More precisely, usually equations (\ref{gen-sys}) have
infinitely many solutions. The main reason is that we fix form of
the Hamiltonian $H$ (\ref{nat-ham}), but an admissible set of the
St\"ackel vectors $s$ in (\ref{st-eq}) remains undefined and,
therefore, unbounded set. So, in order to solve equations
(\ref{gen-sys}) in practice we have to narrow a class of the
separated equations (\ref{st-eq}) once more.

Below we will determine a class of the separated equations by using
the St\"ackel matrix. Recall the St\"ackel matrix is a $n\times n$
block of the transpose Brill-Noether matrix, which is a differential
of the Abel-Jacobi map associated with a product $\mathcal
F(\alpha)$ of the algebraic curves $\mathcal C_i$ (\ref{sep-eq}),
see \cite{ts99d} and references within. There is one to one
correspondence between $j$-th rows of the St\"ackel matrix $S$ and a
basis of the differentials on the  $j$-th curve $\mathcal C_j$.

Let us consider uniform St\"ackel systems \cite{ts99d} for which the
Lagrangian submanifold $\mathcal F(\alpha)$ is the $n$-th symmetric
product of a hyperelliptic curve $\mathcal C$. For the standard
basis of the holomorphic differentials on $\mathcal C$ associated
with the Veronese map the corresponding St\"ackel matrix is equal to
\bq\label{st-mat}
S=\left(%
\begin{matrix}
  \lambda_1^{n-1} & \ldots & \lambda_1 & 1 \\
    \vdots & \ddots & \vdots & \vdots \\
  \lambda_n^{n-1} & \ldots & \lambda_n & 1
\end{matrix}%
\right),\quad\Leftrightarrow\quad \Phi_i=\sum_{k=0}^{n-1}
\alpha_k\lambda_i^k=s_i(\mu_i,\lambda_i)\,.
\eq
It is easy to prove that the control matrix
\bq\label{f1-pedr}
F=S^{-1}\Lambda S=\left(%
\begin{matrix}
  \sigma_1 & 1 & 0 &\cdots & 0 \\
  \sigma_2 & 0 & 1 & \cdots &0 \\
  \vdots &  & \cdots &0 & 1 \\
  \sigma_n & 0 & \cdots & &0
\end{matrix}%
\right)\,,
\eq
fulfills (\ref{42-pedr}) \cite{fmp01,fp02}. Here  $\sigma_j$ are
elementary symmetric polynomials on eigenvalues $\lambda_j$ of
tensor $N$
\bq\label{har-pol-pedr}
\Delta(\lambda)=\lambda^n-\sigma_1\lambda^{n-1}-\sigma_2\lambda^{n-2}-\cdots-\sigma_n\,.
\eq
Solving the separated equations (\ref{st-eq}) with respect to
$\alpha_1,\ldots,\alpha_n $ one gets the  integrals of motion $H_i$
as functions on the separated variables
\bq\label{ham-st}
H_i=\sum_{j=1}^n\dfrac{\partial\sigma_j}{\partial
\lambda_j}\dfrac{s_i(\mu_i,\lambda_i)}{\prod_{i\neq k}
(\lambda_i-\lambda_k)}\,.
\eq
The corresponding vector fields $X_{H_i}$ are tangent to a
bi-Lagrangian foliation, but they are not, in general,
bi-Hamiltonian. These vector fields are  Pfaffian
quasi-bi-hamiltonian fields in the terminology of \cite{mt97}.

\begin{rem}
The St\"ackel matrix $S$ (\ref{st-mat})  is one of the most studied
matrices, which appears very often in various applications
\cite{ben93,fmp01,fp02,imm00,ts99d}. But we have to keep in mind
that there are many other St\"ackel matrices associated with
hyperelliptic or non-hyperelliptic curves.
\end{rem}

\begin{exam} Let us consider natural integrable systems on the plane
separable in the polar coordinates $(r,\phi,p_r,p_\phi)$. The
corresponding second Poisson tensor $P^{\,\prime}$ is degenerate
\bq\label{P-pol}
P^{\,\prime}=\left(%
\begin{matrix}
 0 & 0 & r^2 &0 \\
 * & 0 & 0& 0 \\
 * & * & 0 & 0 \\
 * & * & * & 0
\end{matrix}%
\right)\,.
\eq
Operator $N$ (\ref{recc-op})  has two different eigenvalues
$\lambda_1=r^2$ and $\lambda_2=0$.  Integrals of motion
\bq\label{h-phi}
H_1=p_r^2+\dfrac{p_\phi^2}{r^2}+V(r),\qquad H_2=p_\phi\,
\eq
give rise to the following control matrix $F$ and the separated
equations $\Phi_{1,2}$
\[
F=\left(\begin{array}{cc} r^2&0\\0&0\end{array}\right)\,,\qquad
\left\{\begin{array}{l}
\Phi_1=\quad p_r^2+\dfrac{\alpha_2^2}{r^2}+V(r)-\alpha_1=0,\\
\Phi_2=\quad p_\phi-\alpha_2=0\,. \end{array}\right.
\]
The corresponding St\"ackel matrix $S$ is non-homogeneous in
contrast with (\ref{st-mat}).  If $V(r)$ is polynomial then the
separated equations $\Phi_1=0$ determines hyperelliptic curve,
whereas curve associated with ${\Phi}_2=0$ is non-hyperelliptic.
 \end{exam}

Summing up, the separation of variables theory for $\omega N$
manifolds outlined above provides intrinsic and algorithmic recipe
to construction of the natural integrable systems of the St\"ackel
type.

The algorithm consists of solution of the extended system of
equations (\ref{jac1}-\ref{jac2}) and (\ref{F-pedr}) with some fixed
 St\"ackel matrix $S$. These equations contain the
following unknown functions $P^{\,\prime}_{ij}$, $V$ and
$H_2,\ldots,H_n$. Solving this overdetermined system of equations in
physical variables we get a Poisson pencil, recursion operator,
Darboux-\Nij\ coordinates and the corresponding natural integrable
system simultaneously.

\begin{rem}
In the proposed algorithm we have to substitute some concrete
St\"ackel matrix $S$ into the equations (\ref{F-pedr}) and then to
solve these equations. In general matrix $S$ defines only first part
of the separated equations (\ref{st-eq}) whereas the St\"ackel
vector $s$ remains arbitrary. However if we use the  St\"ackel
matrices (\ref{st-mat}) associated with the Abel-Jacoby map on the
hyperelliptic curves, we implicitly fix the St\"ackel vector $s$ as
well, because the separated equation has to generate namely
hyperelliptic curve.
\end{rem}

%%%%%%%%%%%%%%%%%%%%%%%%%%%%%%%%%%%%%%%%%%%%%%%%%%%%%%%%%%%%%%%%%%%%%%%%%%%%

\section{Bi-hamiltonian systems on the plane}
\setcounter{equation}{0} In this section we substitute natural
Hamiltonian   $H$ (\ref{nat-ham}) and some polynomial anzats for the
Poisson tensor $P^{\,\prime}$ in
(\ref{jac1},\ref{jac2},\ref{F-pedr}) and describe nontrivial
physical solution of these equations by $n=2$.

\begin{defi}
Solution of the equations (\ref{jac1},\ref{jac2},\ref{F-pedr}) is
nontrivial, if  the functions $H_i$ are functionally independent
integrals of motion, $dH_i\neq 0$ for any $i$,  and potential $V(q)$
is some real function, $V(q)\neq$const.
\end{defi}
Moreover, throughout the rest of the section we  consider  only
physical solutions for which all the integrals of motion are
polynomials in momenta.
\begin{exam} Substituting an independent on momenta  tensor $P^{\,\prime}$
 and the following Hamiltonian $H_1=H=p_1p_2+V(q_1,q_2)$
into (\ref{jac1},\ref{jac2},\ref{F-pedr}) we can get the following
solution
\[
P^{\,\prime}=\left(%
\begin{matrix}
 0 & 0 & 0 & 0 \\
 * & 0 & f(q_1)& 0 \\
 * & * & 0 & 0 \\
 * & * & * & 0
\end{matrix}%
\right)\,\qquad
\begin{array}{l}
  H_1=p_1p_2+\frac{q_2}{4C_1q_1}+f(q_1),\\
   \\
  H_2=\sqrt{q_1}\,\exp(C_1\,p_2^2)\,.
\end{array}
\]
It is example of the  non physical solutions which will not
considered below.
\end{exam}

Of course, integrable systems with two-degrees of freedom are one of
the well studied integrable systems and, therefore, we can only
obtain some new tensors $N$ for known integrable systems listed in
\cite{hiet87}. Nevertheless below we present two new integrable
systems on the plane with pseudo-Euclidean metric as well.

\subsection{Linear Poisson tensors and the L-systems}
Let  $Q$ be an $n$-dimensional Riemannian manifold endowed with a
conformal Killing  tensor $L$ of gradient type with vanishing \Nij\
torsion \cite{ben93}. According to \cite{imm00}  its cotangent
bundle $M=T^*Q$ is an $\omega N$ manifold, whose recursion operator
\bq\label{N-st}
 N\,\dfrac{\partial }{\partial  q^k}=\sum_{i=1}^n L_k^i
\dfrac{\partial }{\partial  q^i}+\sum_{ij} p_j\left(\dfrac{\partial
L^j_i}{\partial q^k}-\dfrac{\partial  L^j_k}{\partial
q^i}\right)\dfrac{\partial }{\partial  p_i},\qquad
N\,\dfrac{\partial }{\partial  p_k}=\sum_{i=1}^n L^k_i
\dfrac{\partial }{\partial p_i} \
\eq
is a complete lifting of operator $L$. In this case the Poisson
tensor
\bq\label{P-mag}
P^{\,\prime}=N\,P=\left(%
\begin{matrix}
 0 & -L^j_i \\
L^j_i \quad & \left(\dfrac{\partial L^k_j}{\partial
q_i}-\dfrac{\partial L^k_i}{\partial q_j} \right)p_k
\end{matrix}%
\right)
 \eq
is a linear matrix polynomial in momenta.

Substituting a Poisson tensor $P^{\,\prime}$ (\ref{P-mag}) and a
natural Hamiltonian $H_1=T+V$ (\ref{nat-ham}) into the system of
equations (\ref{F-pedr}) and expanding it in powers of the momenta
we can see that (\ref{gen-sys}) splits into the  equations on  $L$
and on $V$
\bq
d(\mathcal L_{X_T}\,\theta^{\,\prime}-T d\sigma_1)=0,\qquad
d(\mathcal L_{X_V}\,\theta^{\,\prime}-V d\sigma_1)=0\,,
\label{33-pedr}
\eq
and the cyclic recurrence relations for other integrals of motion
\bq\label{int-magri-n}
N^*\,d{H}_i=d{H}_{i+1}+\sigma_id{H}_1\,,\qquad i=2,\ldots,n,
\eq
which we have to solve starting with $i=n$ and $H_{n+1}=0$, see e.g.
\cite{bfp03,imm00,ts05a}.

Here $\mathcal L$ is a Lie derivative along the vector field  $X_T$
or $X_V$ and $\theta^{\,\prime}=\sum_{i,j=1}^n L^i_j\,p_idq^j$ is a
deformation of the  Liouville form $\theta=\sum p_j dq^j$ for any
set of fibered coordinates $(p,q)$.

In \cite{ts05a,ts05b} the computer program for finding of the
L-tensors and the associated integrable systems was constructed.
This software allows to solve equations (\ref{33-pedr}) and
(\ref{int-magri-n}) on the Riemannian manifolds with a
positive-definite metric.

The eigenvalues on  $N$ (\ref{N-st}) coincide with the eigenvalues
$\lambda_i(q_1,\ldots,q_n)$ of the operator $L$ defined on the
configuration space $Q$. As a consequence,  functions $\sigma_i$
depend on  coordinates $q$ only and, according to
(\ref{int-magri-n}), for the natural integrable systems with
$H_1=T+V$ (\ref{nat-ham}) all the integrals of motion
$H_2,\ldots\,H_n$ are quadratic polynomials in the momenta
\cite{ben93,imm00}.

The associated with $N$ (\ref{N-st}) Darboux-\Nij\ coordinates are
related with initial physical variables $(p,q)$ by the point
canonical transformations. Indeed in this case Darboux-\Nij\
coordinates coincide with well-studied orthogonal curvilinear
coordinates on the Riemannian manifolds.

\begin{exam}
On the plane $Q=\mathbb R^2$ there are four $L$-tensors, which give
rise to the natural integrable systems \cite{ben93}. The
corresponding Poisson tensors $P^{\,\prime}$ (\ref{kepl-tenz})
generate elliptic and parabolic coordinates, tensor $P^{\,\prime}$
(\ref{P-pol}) gives rise to the polar coordinates and numerical
skewsymmetric tensor generates cartesian web on the plane
\cite{ben93}.
\end{exam}

\begin{rem}
On the plane $Q=\mathbb R^2$ there are many other solutions of the
equations (\ref{jac1},\ref{jac2}) in the form (\ref{P-mag})
\bq\label{P-lin1}
P^{\,\prime}=\left(%
\begin{matrix}
A p & B \\
-B \quad & C p
\end{matrix}%
\right),
 \eq
where entries of the matrices $A,B,C$ depend on coordinates $q$.
However if $H_1=T+V$ is a natural Hamiltonian (\ref{nat-ham}) there
are only four nontrivial solutions of the extended system of
equations (\ref{jac1},\ref{jac2},\ref{F-pedr}) associated with the
four known coordinate systems.
\end{rem}

\subsection{Linear Poisson tensors and systems with cubic integrals of motion}
In this section we consider the following ansatz for the Poisson
tensor
\bq\label{lin-gen}
P^{\,\prime}=\left(%
\begin{matrix}
 0 & h(q) & c_1p_1+c_2p_2 & c_3p_3+c_4p_4 \\
 * & 0 & c_5p_1+c_6p_2& c_7p_1+c_8p_2 \\
 * & * & 0 & f(q) \\
 * & * & * & 0
\end{matrix}%
\right)\,,\qquad c_k\in \mathbb R.
\eq
Here $c_k$ are undetermined coefficients and $f,h$ are some
functions on coordinates $q$.

Substituting $P^{\,\prime}$ (\ref{lin-gen}) into the Jacobi
equations (\ref{jac1},\ref{jac2}) we can extract equations for the
function $h$. Solving these equations it is easy to prove that
either function $h(q)$ is a constant $h(q)=const$, or it is a linear
function of coordinates $h(q)=aq_1+bq_2$.

\begin{prop}
If the Poisson tensor $P^{\,\prime}$ has the form (\ref{lin-gen})
and integrals of motion \[H_1=p_1^2+p_2^2+V(q_1,q_2),\qquad
H_2=H_2(p_1,p_2,q_1,q_2),\] satisfy to the  Lenard-Magri recurrence
relations (\ref{bi-chain}) then the extended system of equations
(\ref{jac1},\ref{jac2},\ref{F-pedr})  with control matrix $F$
(\ref{f-num}) has only one nontrivial solution up to canonical
transformations
\bq\label{lin-toda}
P^{\,\prime}=\left(\begin{matrix} 0& a& p_1& 0\\
*& 0& 0& p_2\\
*& 0& 0& b\, e^{\frac{q_2-q_1}{a}}\\
*&*&*& 0\end{matrix}\right),\qquad
H_1=p_1^2+p_2^2+2a\,b\,e^{\frac{q_2-q_1}{a}}\,.
\eq
The corresponding second integral of motion
$H_2=(p_1+p_2)H_1-\frac13(p_1+p_2)^3$  is formally a third order
polynomial in the momenta.
\end{prop}
This integrable system coincides with open Toda lattice of $\mathcal
A_2$ type.

\begin{rem}\label{rem-toda1}
From the Lenard-Magri relations (\ref{bi-chain}-\ref{F-pedr}) with
control matrix $F$ (\ref{f-num}) it follows that
\bq \label{b2-toda}
\begin{array}{cc}
{N^*}dH_1=dH_2, \qquad& {N^*}dH_2=\sigma_2dH_1+\sigma_1dH_2,\\
\\
{N^*}^2dH_1=dH_2^{(2)}, \qquad&
{N^*}^2dH_2^{(2)}=\sigma_2^{(2)}dH_1+\sigma_1^{(2)}dH_2^{(2)},\\
\vdots&\vdots
\end{array}
\eq
where $dH_2^{(2)}=\sigma_2dH_1+\sigma_2dH_2$ and $\sigma_i^{(2)}$
are symmetric polynomials associated with ${N^*}^2$. Similar
relations (\ref{b2-toda}) may be constructed only for the
${N^*}^{2^m}$.

In our case (\ref{lin-toda}) $H_2^{(2)}=(p_1p_2
-ab\,\exp((q_2-q_1)/a)\, )^2-\frac{1}{2}H_1^2$ is a fourth order
polynomial in the momenta. Below we will use deformations of the
operator $N^2$ in order to construct open Toda lattices of $\mathcal
BC_2$ and $\mathcal D_2$ type.  To get the Toda lattice of $\mathcal
G_2$ type  we can use deformations of the operator $N^4$.
\end{rem}

\begin{rem}\label{rem-toda2}
The   Poisson tensor $P^{\,\prime}$ (\ref{lin-toda}) admits a
natural multi-dimensional generalization
\bq\label{toda-gen}
P^{\,\prime}=\sum_{i=1}^{n-1}
2e^{2(q_i-q_{i+1})}\dfrac{\partial}{\partial
p_{i+1}}\wedge\dfrac{\partial}{\partial p_{i}} +\sum_{i=1}^n
p_i\dfrac{\partial}{\partial q_{i}}\wedge\dfrac{\partial}{\partial
p_{i}}+\dfrac12\sum_{i<j}^n \dfrac{\partial}{\partial
q_{j}}\wedge\dfrac{\partial}{\partial q_{i}}
\eq
 which was considered in \cite{fern93}.
 The corresponding open Toda lattice of $\mathcal A_n$ type  are one
of the  well-studied bi-hamiltonian systems.
\end{rem}

Throughout the rest of the section we will use the
 matrix $F$ (\ref{f1-pedr}) associated with the homogeneous St\"ackel
 matrix $S$ (\ref{st-mat}) of the Veronese type.

\begin{prop}\label{prop-lin}
If the Poisson tensor $P^{\,\prime}$ has the form (\ref{lin-gen})
and \[H_1=p_1^2+p_2^2+V(q_1,q_2),\qquad H_2=H_2(p_1,p_2,q_1,q_2),\]
then  equations (\ref{jac1},\ref{jac2},\ref{F-pedr})  with matrix
$F$ (\ref{f1-pedr}) has only four nontrivial solutions up to
canonical transformations
\ben
&P^{\,\prime}=\left(\begin{matrix} 0& -2& p_2& 3 p_1\\
*& 0& p_1& p_2\\
*& *& 0& b e^{q_1}\\  *& *& -b e^{q_1}& 0
\end{matrix}\right)\,,\quad
&V=-\frac{2b}{3}\,e^{q_1}+C_1e^{\left(-\frac{q_1}2+\frac{\sqrt{3}q_2}{2}\right)}
+C_2e^{\left(-\frac{q_1}2-\frac{\sqrt{3}q_2}{2}\right)} \,,
\nn\\
\nn\\
&P^{\,\prime}=\left(\begin{matrix} 0& q_1& -\frac23 p_2& 0\\
*& 0&\frac13 p_1& -\frac23 p_2\\
*&*& 0& b q_1^{1/3}\\  *& *& *& 0\end{matrix}\right),\quad
&V=\dfrac{b(3q_1^2+4q_2^2)+C_1q_2+C_2}{4q_1^{2/3}}\,,
\nn\\
\nn\\
&P^{\,\prime}=\left(\begin{matrix} 0& q_1& 0& 0\\
*& 0& p_1& -2 p_2\\  *& *& 0& b q_1\\ *& *&*&
0\end{matrix}\right),\quad
&V=b\left(\dfrac{q_1^2}4+q_2^2\right)+C_1q_2+\frac{C_2}{q_1^2}\,,
\nn\\
\nn\\
&P^{\,\prime}=\left(\begin{matrix} 0& q_1& 0& 4 p_1\\
*& 0& p_1& 2 p_2\\
*& *& 0& \dfrac{b}{q_1^3}\\  *& *& *& 0\end{matrix}\right),\quad
&V=\frac{b}{4q_1^2}+C_1q_2^2+C_2q_2-C_3q_1^2\,.
\nn
\en
Here $C_j$ are arbitrary constants, which appear by integration of
differential equations.
\end{prop}
The corresponding integrals of motion are the third order
polynomials in the momenta $p$, which may be found in \cite{hiet87}.
The first solution is a periodic Toda lattice of $\mathcal A_3$ type
in the center of mass system \cite{hiet87}. The second solution is a
Holt system \cite{hiet87,ho82}. Two remaining solutions may be
considered as integrable systems of Calogero type.

In fact we proved that all these systems are the uniform St\"ackel
systems with the homogeneous St\"ackel matrix $S$ (\ref{st-mat}). It
is easy to check that the corresponding separated equations give
rise to the hyperelliptic curves.

\begin{rem} There are many other solutions of the
equations (\ref{jac1},\ref{jac2}) if we suppose that coefficients
$c_k$ in  $P^{\,\prime}$ (\ref{lin-gen}) are some functions of
coordinates $q$. However if $H_1=T+V$ is a natural Hamiltonian
(\ref{nat-ham}) there are only four nontrivial solutions of the
extended system of equations (\ref{jac1},\ref{jac2},\ref{F-pedr}).
\end{rem}

We fix the form of the desired Poisson tensor $P^{\,\prime}$
(\ref{lin-gen}) and, therefore, in the equations (\ref{F-pedr}) we
could use the natural Hamilton function defined up to canonical
transforms
\bq\label{gen-nat}
H=\widetilde{T}+V(q),\qquad \widetilde{T}=\sum_{i,j=1}^n
b_{ij}(q)p_ip_j+\sum_{k=1}^n b_k(q) p_k\,.
\eq
Here $b_{ij}(q)$ and  $b_k(q)$ are functions of coordinates $q$,
such that kinetic energy $\widetilde{T}$ may be reduced to the
canonical form (\ref{nat-ham}) by  point transformations $q\to
q'=Aq+b$ and canonical shifts of the momenta $p_i\to p_i+b_i(q_i)$.

As an example we consider natural Hamiltonians with pseudo-Euclidean
metric \cite{dr35} and reproduce one  Drach system which is not
superintegrable St\"ackel system with quadratic integrals of motion
\cite{ts00f}. Only for this Drach system the separated variables was
unknown.

\begin{prop}
If the Poisson tensor $P^{\,\prime}$ has the form (\ref{lin-gen})
and  $H_1=p_1p_2+V(q_1,q_2)$ then one of the solutions of the
equations  (\ref{jac1},\ref{jac2},\ref{F-pedr})  reads
\[
P'=\left(\begin{matrix}0& q_1& 0& 0\\
                              *& 0& p_1+a p_2&-\frac{p_2}{2}\\
                              *&*& 0& \frac{b}{\sqrt{q_1}}\\
                              *&*&*&0\end{matrix}\right),\qquad
H_1=p_1p_2-2b\dfrac{2aq_1-q_2}{\sqrt{q_1}}+\dfrac{C_1}{\sqrt{q_1}}+C_2\left(q_2-\frac{2a}{3}q_1\right)\,,
\]
The second integral of motion $H_2$ is a third order polynomial in
 momenta. The corresponding separated variables
are the special Darboux-\Nij\ variables
\[
\lambda_{1,2}=-\frac{p_2}{4}\pm\frac{\sqrt{p_2^2+16\,b\,\sqrt{\,q_1\,}}}{4}\,
\]
and the separated equations give rise to hyperelliptic curve.
\end{prop}
\begin{rem}
In this section we identify the first integral $H_1$ from the \Nij\
chain (\ref{F-pedr}) with the natural Hamiltonian $H=T+V$. Of
course, we can choose any integral $H_k$ from the chain as the
natural Hamilton function.
\end{rem}

\subsection{The Poisson tensors second order in the momenta}
In this section we consider the following ansatz for the Poisson
tensor
\bq\label{gen-2}
P^{\,\prime}=\left(\begin{array}{rrrr}
0&  h_1(q)p_1+h_2(q)p_2& \sum c^{1}_{ij}p_ip_j+\mathrm g_1(q)&\sum c^{2}_{ij}p_ip_j+\mathrm g_2(q)\\
*& 0&\sum c^{3}_{ij}p_ip_j+\mathrm g_3(q)& \sum c^{4}_{ij}p_ip_j+\mathrm g_4(q)\\
*& * & 0 & f_1(q)p_1+f_2(q)p_2 \\
*& * & * & 0\end{array}\right)\,,\qquad c_k\in \mathbb R.
\eq
Here $f_i,\mathrm g_i,h_i$ are some functions on coordinates
 $q$ and $c^{k}_{ij}$ are undetermined numerical
coefficients.

As above there are solutions with  potentials $V$ separable in
physical coordinates $q_{1,2}$,  $\frac{\partial ^2 V}{\partial
q_1\partial q_2}=0$, and solutions with formally fourth order
integrals of motion for which $H_2$ consists of $H_1$ and another
independent quadratic polynomial $\widetilde{H}_2$, for instance
$H_2=\alpha H_1^2+\widetilde{H}_2$ or $H_2=H_1\widetilde{H}_2$. In
contrast with the previous section we will not present  such
solutions below.

\begin{exam}
By using tensor $P^{\,\prime}$ (\ref{gen-2}) we can obtain
integrable systems with quadratic integrals of motion. For instance
one of the solutions of the equations
(\ref{jac1},\ref{jac2},\ref{F-pedr}) with the control matrix $F$
(\ref{f1-pedr}) is equal to
\[
H_1=p_1^2+p_2^2+\frac{C_3}{2}\Bigl(q_1^2+q_2^2\Bigr)
-\frac{C_1-C_2}{2(q_1+q_2)^2}-\frac{C_1+C_2}{2(q_1-q_2)^2}
\]
and
\bq\label{qInt}
P^{\,\prime}=\left(\begin{array}{rrrr} 0& p_1 q_2-p_2 q_1&
-p_1^2+\mathrm
g_1(q)& -p_2p_1+\mathrm g_2(q)\\
*& 0& -p_2 p_1+\mathrm g_3(q)& -p_2^2+\mathrm g_4(q)\\
*& *& 0& p_1 f_1(q)+p_2 f_2(q)\\
*& *& *& 0
\end{array}\right)
\eq
where
\[
\begin{array}{lll}
\mathrm
g_1=\frac{q_1(C_1q_1^3+3C_2q_1^2q_2+3C_1q_1q_2^2+C_2q_2^3)}{(q_1-q_2)^3(q_1+q_2)^3},\qquad
& \mathrm g_3=q_2q_1^{-1}\,\mathrm g_1(q)\,,\qquad&
f_2=q_1^{-1}\mathrm g_1(q)
\\
\\
\mathrm
g_2=\frac{q_1(C_2q_1^3+3C_1q_1^2q_2+3C_2q_1q_2^2+C_1q_2^3)}{(q_1-q_2)^3(q_1+q_2)^3}\,,
\qquad&\mathrm g_4=q_2q_1^{-1}\,\mathrm g_2(q)\,,\qquad &
f_1=q_1^{-1}\mathrm g_2(q)\,.
\end{array}
\]
The corresponding second integral of motion
$H_2=\widetilde{H}_2-\frac14 H_1^2$ is a fourth order polynomial in
 momenta, which may be reduced to a quadratic integral.

The associated with $P^{\,\prime}$ (\ref{qInt}) solution of the
equations (\ref{jac1},\ref{jac2},\ref{F-pedr}) with another control
matrix $F$ (\ref{f-num}) has a form
\[H_1=
p_1^2+p_2^2-\frac{C_1(q_1^2+q_2^2)}{(q_1+q_2)^2(q_1-q_2)^2}-\frac{2C_2q_1q_2}{(q_1+q_2)^2(q_1-q_2)^2}\,.
\]
As above  second integral of motion is reduced to a quadratic
polynomial in the momenta.
\end{exam}

\par\noindent
\textbf{Integrals of motion fourth order in the momenta.}
\vskip0.1truecm
\par\noindent
Below we consider integrable systems for which second integrals of
motion are the fourth order polynomials in the momenta, which can
not be reduced to quadratic polynomials.

Substituting $P^{\,\prime}$ (\ref{gen-2}) into the equations
(\ref{jac1},\ref{jac2}) we can extract equations for the functions
$h_i$. Solving these equations we get that either functions $h_i$
are constant $h_i=const$, or  they are linear functions of
coordinates  $h_i=a_iq_1+b_iq_2$.

At $h_i=const$ we consider only the Lenard-Magri recurrence
relations. Recall that in this case the control matrix $F$ in
(\ref{F-pedr}) has a special form (\ref{f-num}).

\begin{prop}
If the Poisson tensor $P^{\,\prime}$ has the form (\ref{gen-2})
there are only two solutions of the extended system of equations
(\ref{jac1},\ref{jac2},\ref{F-pedr}) with a control matrix $F$
(\ref{f-num}) and  a natural Hamilton function
$H_1=p_1^2+p_2^2+V(q)$.

The corresponding Poisson tensors are distinguished by functions
$f_i$ and $\mathrm g_i$ only
\bq\label{exp-Toda} P^{\,\prime}=\left(\begin{array}{cccc}
0 & 2ap_1+2bp_2& p_1^2+\mathrm g_1(q)& \mathrm g_2(q)\\
* & 0& \mathrm g_3(q)& p_2^2+\mathrm g_4(q)\\
* & * & 0 & f_1(q)p_1+f_2(q)p_2\\
* & * & * & 0
\end{array}\right)\,.
\eq
If $a\neq 0$ and $b\neq 0$ then the Hamiltonian reads as
\[
H_1=p_1^2+p_2^2+(a+b)C_2\exp\left(\dfrac{q_2-q_1}{a+b}\right)
               -(a-b)C_1\exp\left(\dfrac{q_2+q_1}{a-b}\right)\,,
\]
whereas  functions $f_i$ and $\mathrm g_i$ in (\ref{exp-Toda}) are
equal to
\[
\begin{array}{lll}
f_1=C_2e^{\frac{q_2-q1}{a+b}}-C_1e^{\frac{q_1+q_2}{a-b}},\qquad& \mathrm g_1=2bf_2,\qquad& \mathrm g_2=bf_1-af_2\\
f_2=C_2e^{\frac{q_2-q1}{a+b}}+C_1e^{\frac{q_1+q_2}{a-b}},\qquad&
\mathrm g_3=-\mathrm g_2,\quad&\mathrm  g_4=2af_1\,.
\end{array}
\]
If $a=0$ or $b=0$, then this is another nontrivial solution. At
$a=0$ and  $b=1$ the corresponding Hamiltonian reads as
\bq\label{toda-Dn}
H_1=p_1^2+p_2^2+2C_1e^{-q_1-q_2}+2C_2e^{q_2-q_1}-
\frac{C_3e^{2q_2}}{2(C_2e^{2q_2}-C_1)^2}
-\frac{C_4e^{q_2}(C_2e^{2q_2}+C_1)}{(C_2e^{2q_2}-C_1)^2}
\eq
whereas functions $f_i$ and $\mathrm g_i$ in (\ref{exp-Toda}) are
equal to
\[
\begin{array}{lll}
f_1=(C_2e^{q_2}-C_1e^{-q_2})e^{-q_1},\quad& \mathrm g_1=2f_2,\quad& \mathrm g_2=f_1+\frac{d \mathrm g_4}{dq_2}\\
\\
f_2=(C_2e^{q_2}+C_1e^{-q_2})e^{-q_1},\quad& \mathrm g_3=f_1,\quad&
\mathrm
g_4=\frac{C_3e^{2q_2}}{(C_2e^{2q_2}-C_1)^2}+\frac{C_4e^{q_2}(C_2e^{2q_2}+C_1)}{(C_2e^{2q_2}-C_1)^2}\,.
\end{array}
\]
In the both cases the second integrals of motion are the fourth
order polynomials in the momenta.
\end{prop}
The first solution is an open Toda lattice of $\mathcal D_2$ type.
If $C_3=C_4=0$ the second Hamiltonian $H_1$ (\ref{toda-Dn})
describes open Toda lattice  of $\mathcal D_2$ type. If $C_1=0$ or
$C_2=0$ the Hamiltonian $H_1$ (\ref{toda-Dn}) describes also an open
Toda lattice associated with a root system $\mathcal{BC}_2$. If
$C_1=C_2$ we obtain the Hamiltonian  $H_1$ (\ref{toda-Dn}) of the
generalized Toda lattice, which was constructed by Inozemtsev
\cite{in89}.

\begin{rem}
The quadratic matrix polynomial $P^{\,\prime}$ (\ref{exp-Toda}) may
be considered as deformation of the following  product of  linear
Poisson tensor (\ref{lin-toda}) (see Remark \ref{rem-toda1})
\bq \label{exp-Toda2}
P^{\,\prime}=P^{\,\prime}_{lin}P^{-1}P^{\,\prime}_{lin}+ K_{lin}\,,
\eq where $K_{lin}$ is a very
simple linear matrix polynomial. Therefore,  substituting the
multi-dimensional linear Poisson tensor (\ref{toda-gen}) into
(\ref{exp-Toda2}) and  adding the similar term $K_{lin}$ one gets a
natural multi-dimensional generalization of $P^{\,\prime}$
(\ref{exp-Toda}), which was proposed in \cite{dm04} for the
 open Toda lattices at $C_3=C_4=0$ and at $C_1=0$ ($C_2=0$).
\end{rem}

Now let us consider the second case at  $h_i=a_iq_1+b_iq_2$.
\begin{prop}
If the Poisson tensor $P^{\,\prime}$ has the form (\ref{gen-2})
there are  only two explicit nontrivial solutions of the equations
(\ref{jac1},\ref{jac2},\ref{F-pedr}) with control matrix $F$
(\ref{f1-pedr}) and  a natural Hamilton function
$H_1=p_1^2+p_2^2+V(q)$.

The first solution is the  Holt-type system for which
\bq\label{Holt-9}
H_1=p_1^2+p_2^2-\dfrac{C_1(2q_1^2+9q_2^2)}{6q_2^{2/3}}+\dfrac{C_2}{q_1^2}
+C_3q_2^{2/3}-\dfrac{C_4}{q_2^{2/3}}
\eq
and
\bq\label{rad-3}
P^{\,\prime}=\left(\begin{matrix}
 0\quad&  3p_1q_2\quad& p_1^2+\frac{C_2}{q_1^2}\quad&p_1p_2+C_1q_1q_2^{1/3}\\
*& 0& -\frac{3C_2q_2}{q_1^3}\quad& p_1^2-3C_1q_2^{4/3}+\frac{C_2}{q_1^2}+2C_3q_2^{2/3}\\
*& * & 0 & -C_1p_1q_2^{1/3}-\frac{C_2p_2}{q_1^3}\\
*& * & * & 0\end{matrix}\right)\,.
\eq
The second solution is the system with fourth order potential for
which
\bq\label{4-pot}
H_1=p_1^2+p_2^2-C_1\left(4q_1^4+3q_1^2q_2^2+\frac{q_2^4}2\right)+C_2\left(q_1^2+\frac{q_2^2}4\right)
-\frac{C_3}{q_1^2}-\frac{C_4}{q_2^2}
\eq
and
\bq\label{rad-2}
P^{\,\prime}=\left(\begin{matrix}
0&  p_1q_2& p_1^2+\mathrm g_1(q)&p_1p_2+\mathrm g_2(q)\\
*& 0& \mathrm g_3(q)& p_1^2+2p_2^2+\mathrm g_4(q)\\
*& * & 0 & p_1f_1(q)+p_2f_2(q)\\
*& * & * & 0\end{matrix}\right)\,,
\eq
where
\[
\begin{array}{ll}
  f_1=-C_1q_2^3,\quad&
  f_2=2C_1q_1(4q_1^2+3q_2^2)-C_2q_1-\dfrac{C_3}{q_1^3},\\
  \mathrm g_2=C_1q_1q_2^3,\quad&
  \mathrm g_1=-4C_1q_1^4+C_2q_1^2-\dfrac{C_3}{q_1^2},\\
   \mathrm g_3=4\mathrm g_2+q_2f_2,\quad&
  \mathrm g_4=-C_1(4q_1^4+6q_1^2q_2^2+q_2^4)+C_2\left(q_1^2+\dfrac{q_2^2}2\right)-\dfrac{C_3}{q_1^2}
  \end{array}
\]
The second integrals of motion are the fourth order polynomials in
the momenta \cite{hiet87}.
\end{prop}
According to \cite{ts96a}, in the second case we can make a special
contraction of variables and obtain the Henon-Heiles system with the
cubic potential
\bq\label{3-pot}
H_1= p_1^2+p_2^2+\frac{1}{3}q_1(16q_1^2+3q_2^2)+\frac{C_1}{q_2^2}\,.
\eq
The corresponding Poisson tensor $P^{\,\prime}$ is equal to
\bq\label{hen-16}
P^{\,\prime}=\left(\begin{matrix}0& q_2p_1& p_1^2+\frac{16}{3}q_1^3&
p_1p_2-\frac{1}{6}q_2^3\\
*& 0& 8q_2q_1^2+\frac{1}{3}q_2^3&
p_1^2+2p_2^2+2q_1q_2^2+\frac{16}{3}q_1^3\\  *& *& 0&
-p_2(8q_1^2+q_2^2)\\ *& *& *& 0\end{matrix}\right)\,.
\eq
\begin{rem}
The eigenvalues of the operators $N$ associated with the Poisson
tensors (\ref{rad-2}) and (\ref{hen-16}) are integrals of motion
$\dot{\lambda}_{1,2}=0$. So, the special Darboux-\Nij\ coordinates
can not be the separated variables for the system with fourth order
potential (\ref{4-pot}) and for the Henon-Heiles system
(\ref{3-pot}). The corresponding separated variables are generic
Darboux-\Nij\ coordinates $(x,y)$ (\ref{xy-pedr}).
\end{rem}

If the Poisson tensor $P^{\,\prime}$ has a form  (\ref{gen-2}) then
there are some  implicit solutions of the equations
(\ref{jac1},\ref{jac2},\ref{F-pedr}) with control matrix $F$
(\ref{f1-pedr}).  In these cases functions $f_i(q),\mathrm g_i(q)$
and integral of motion $H_2(p,q)$ are determined by potential
$V(q_1,q_2)$, which satisfies some special partial differential
equations of the second and of the third order. For instance, one of
the obtained equations has the form
\[
(q_1q_2+C_1)\left(\frac{\partial^2}{\partial
q_1^2}-\frac{\partial^2}{\partial q_2^2}\right)
V+3\left(q_2\frac{\partial}{\partial
q_1}-q_1\frac{\partial}{\partial
q_2}\right)V=(q_1^2-q_2^2+C_2)\frac{\partial^2}{\partial q_1\partial
q_2}V\,.
\]
Similar second order equations were considered in
\cite{hiet87,rw84}. The third order partial differential equations
obtained in this method we could not find in the literature. These
implicit systems will be studied separately.

As above we can substitute into the equations (\ref{F-pedr}) the
Hamilton function $H_1=T+V$ in the form  (\ref{gen-nat}). Here we
present two new integrable systems with pseudo-Euclidean metric.
\begin{prop}
If the Poisson tensor $P^{\,\prime}$ has a form  (\ref{gen-2}) there
are only two explicit solutions of the equations
(\ref{jac1},\ref{jac2},\ref{F-pedr}). The first solution reads as
\[
H_1=p_1p_2+C_1(4q_1^2+5\alpha
q_1q_2+\alpha^2q_2^2)+\dfrac{C_2(q_1+\alpha q_2)}{q_2^{1/3}}+
\dfrac{C_3}{q_2^{2/3}}+\dfrac{C_4}{q_2^{4/3}}\,,\qquad
\alpha\in\mathbb R,
\]
\[
P^{\,\prime}=\left(\begin{smallmatrix} 0& 3q_2p_1\qquad&
p_1^2-6C_1q_2(q_1+\alpha q_2)\qquad& \alpha\,p_1^2-3p_1p_2
-3C_1(q_1+\alpha q_2)(2q_1+\alpha q_2) -\frac{C_2(q_1+\alpha
q_2)}{q_2^{1/3}}+\frac{2C_4}{q_2^{4/3}}
\\
\\
*& 0& 18C_1q_2^2& 9C_1q_2(2q_1+\alpha q_2)+3C_2q_2^{2/3}\\
\\
*& *& 0& \left(3C_1(2q_1-\alpha q_2)+\frac{C_2}{q_2^{1/3}}+\right)p_1+18C_1q_2p_2\\
\\
*& *& *& 0
\end{smallmatrix}\right)
\]
The second solution has the form
\[
H_1=p_1p_2+\left(C_1+\frac{C_2}{\sqrt{q_2}}\right)q_1+C_3q_2+\frac{3C_2C_3}{C_1}\sqrt{q_2}\,,
\]
\[
P^{\,\prime}=\left(\begin{smallmatrix} 0& 4q_2p_1\qquad&
p_1^2+q_1+\alpha\,q_2\qquad&(6C_1+4\alpha)p_1^2-2p_1p_2
+2(C_1+\alpha)q_1+2C_1\alpha\left(\left(\sqrt{q_2}+\frac{C_2}{C_1}\right)^2
-\frac{C_2^2(6C_1+\alpha)}{4C_1^3}\right)
\\
\\
*& 0& 2q_2& p_1^2+q_1+(\alpha+4C_1)q_2+4C_2\sqrt{q_2}\\
\\
*& *& 0& -\frac{C_3}{C_1}p_1+p_2\\
\\
*& *& *& 0
\end{smallmatrix}\right)\,,
\]
where  $\alpha=\frac{C_3-2C_1^2}{C_1}$.
\end{prop}
In the both cases the corresponding second integrals of motion are
the fourth order polynomials in the momenta. Moreover, the special
Darboux-\Nij\ coordinates associated with $N$ are the separated
variables and the separated equations give rise to hyperelliptic
curves.

\subsection{The normalized Poisson tensors}
In this section we consider  the normalized Poisson tensors
\bq\label{P-norm}
{P}^{\,\prime}=\det(\widehat{P}^{\,\prime})^{-1}\,\widehat{P}^{\,\prime},
\eq
where $\widehat{P}^{\,\prime}$ has the form (\ref{gen-2}). Such
tensors induce formal $\omega N$ structure on $M$, but their
eigenvalues $\lambda_i$ will be a priori functionally dependent
functions.

\begin{prop}
For the Poisson tensor $P^{\,\prime}$ (\ref{P-norm}) there are only
two nontrivial solutions of the equations
(\ref{jac1},\ref{jac2},\ref{F-pedr}) with the control matrix $F$
(\ref{f1-pedr}).

The corresponding Poisson tensors are distinguished by functions
$f_i$ and $\mathrm g_i$ only
\bq\label{P-norm2}
\widehat{P}^{\,\prime}=\left(\begin{array}{rrrr}
0& p_1q_2&-p_2^2+\mathrm g_1(q)&p_1p_2+\mathrm g_2(q)\\
*& 0& \mathrm g_3(q)& p_2^2+\mathrm g_4(q)\\
* & * & 0& f_1(q)p_1+f_2(q)p_2\\
* & * & *& 0\end{array}\right)\,.
\eq
First solution corresponds to the Henon-Heiles system for which
\[
H_1=p_1^2+p_2^2-C_1q_1(16q_1^2+3q_2^2)-2C_2q_1,
\]
whereas  functions $f_i$ and $\mathrm g_i$ in (\ref{P-norm2}) are
equal to
\[\begin{array}{lll}
f_1=0,\quad &  \mathrm g_1=3C_1q_1q_2^2,\quad &\mathrm g_2=\frac{C_1}{2} q_2^3,\\
\\
f_2= 3C_1(8q_1^2+q_2^2)+C_2,\quad &\mathrm g_3=4\mathrm g_2-f_2q_2,&
\mathrm g_4=-\mathrm g_1\,.
\end{array}
\]
Second solution corresponds to the system with the fourth order
potential for which
\[H_1=p_1^2+p_2^2+C_1(8q_1^4+6q_1^2q_2^2+q_2^4)+ C_2(4q_1^2+q_2^2)+\dfrac{C_3}{q_1^2},\]
 whereas  functions $f_i$ and $\mathrm g_i$ in (\ref{P-norm2}) are equal to
\[\begin{array}{ll}
f_1=2C_1q_2^3,\quad  \mathrm g_1=-(C_1(q_2^2+6q_1^2)+C_2)q_2^2,\quad &\mathrm g_2=-2C_1q_1q_2^3,\\
\\
f_2=-4C_1q_1(3q_2^2+4q_1^2) -4C_2q_1+\frac{C_3}{q_1^3},\quad
&\mathrm g_3=4\mathrm g_2-f_2q_2,\qquad \mathrm g_4=-\mathrm g_1\,.
\end{array}
\]
The corresponding second integrals of motion are equal   to
\[H_2=\det(\widehat{P}')= \sqrt{p_2^4 +q_1f_1p_1^2+(q_2f_2+\mathrm
g_3)p_1p_2 -2\mathrm g_1p_2^2-\mathrm g_1^2+\mathrm g_2\mathrm
g_3}\,.
\]
\end{prop}
In the both cases eigenvalues of the recursion operator $N$ are
trivial $\lambda_{1,2}=\pm 1$. In order to get the separated
variables we have to put operator $N$  in diagonal form
(\ref{xy-pedr}) by using canonical transformation $z=(q,p)\to
\zeta=(x,y)$.

\begin{prop}
Let $J=\frac{\partial \zeta}{\partial z}$ is the Jacobi matrix
associated with some change of variables $\zeta=\zeta(z)$. Solving
overdetermined system of the partial differential equations
\bq\label{sep-tr}
JPJ^T=P,\qquad JNJ^{-1}=\mbox{\rm diag}(1,-1,1,-1)
\eq
with respect to $\zeta(p,q)$ we reproduce the separated variables
 for the Henon-Heiles system and the separated
variables  for the system with fourth order potential, which were
constructed in \cite{ts96a}.
\end{prop}
The first equation in (\ref{sep-tr}) provides canonicity of
transformation $z\to\zeta$, the second one provides diagonability of
$N$ in $\zeta$ variables. Equations (\ref{sep-tr}) were easy solved
 in the symbolic computational system Maple.

Relations of such transformations, which put operator $N$  in
diagonal form, with the B\"acklund transformations for these systems
will be discussed in forthcoming publications.

%%%%%%%%%%%%%%%%%%%%%%%%%%%%%%%%%%%%%%%%%%%%%%%%%%%%%%%%%%%%%%%%%%%%%%

\section{Appendix: a programm for the search of natural integrable systems}
This paper as well as \cite{ts05a,ts05b} belongs mostly to
computational mathematical physics and, therefore, in this section
we present an implementation of the discussed algorithm made in the
symbolic computational system Maple.

At the first step we need to determine  dimension of the phase space
$M$ and  canonical variables
$q=(q_1,\ldots,q_n)$ and $p=(p_1,\ldots,p_n)$:\\
{\footnotesize
\verb"> restart: with(linalg):"\\
\verb"> n:=2:"\\
\verb"> q:=seq(q||i,i=1.. n):  p:=seq(p||i,i=1.. n):"\\
\verb"> var:=q,p;"\\}
Now we can introduce canonical Poisson tensor $P$ (\ref{can-Poi})\\
{\footnotesize
\verb"> ed:=array(identity, 1.. n,1.. n): z:=array(sparse,1.. n,1.. n):"\\
\verb"> P1:=blockmatrix(2,2,[z,ed,-ed,z]);"\\}
and a linear ansatz for the second Poisson tensor $P^{\,\prime}$
(\ref{lin-gen}) at $h(q)=aq_1$\\
{\footnotesize
\verb"> P2 := array(antisymmetric,1.. 2*n,1.. 2*n):"\\
\verb"> P2[1,2]:=a*q1:        P2[1,3]:=c1*p1+c2*p2: P2[1,4]:=c3*p1+c4*p2:"\\
\verb"> P2[2,3]:=c5*p1+c6*p2: P2[2,4]:=c7*p1+c8*p2: P2[3,4]:=f1(q):"\\}
The Poisson brackets associated with the Poisson tensor
$P^{\,\prime}$
and with the Poisson pencil $P+\lambda P^{\,\prime}$ are defined by the standard rule\\
{\footnotesize
\verb"> P3:=evalm(P1+lambda*P2);"\\
\verb"> PB2 := proc (f, g) options operator, arrow:"\\
\verb">       add(add(P2[i,j]*diff(f,var[i])*diff(g,var[j]),i=1.. 2*n),j=1.. 2*n)"\\
\verb">       end:"\\
\verb"> PB3 := proc (f, g) options operator, arrow:"\\
\verb">       add(add(P3[i,j]*diff(f,var[i])*diff(g,var[j]),i=1.. 2*n),j=1.. 2*n)"\\
\verb">       end:"\\}
The corresponding Jacobi identities (\ref{Jac-id2}) look like \\
{\footnotesize
\verb"> f:=F(var); g:=G(var): h:=H(var):"\\
\verb"> Y2:=simplify(PB2(PB2(f,g),h)+PB2(PB2(g,h),f)+PB2(PB2(h,f),g)):"\\
\verb"> Y3:=simplify(PB3(PB3(f,g),h)+PB3(PB3(g,h),f)+PB3(PB3(h,f),g)):"\\}
It allows us to build up the system of equations (\ref{jac1},\ref{jac2})\\
 {\footnotesize
\verb"> ListEq:=NULL:"\\
\verb"> for i to 2*n do"\\
\verb">   for j to 2*n do"\\
\verb">      for k to 2*n do"\\
\verb">      Eq2:=coeff(coeff(coeff("\\
\verb">      Y2,diff(f,var[i]),1),diff(h,var[j]),1),diff(g,var[k]),1):"\\
\verb">      Eq3:=coeff(coeff(coeff("\\
\verb">      Y3,diff(f,var[i]),1),diff(h,var[j]),1),diff(g,var[k]),1):"\\
\verb">      ListEq:=ListEq,coeffs(collect(Eq2,{p},distributed),{p}),"\\
\verb">              coeffs(collect(Eq3,{p,lambda},distributed),{p,lambda});"\\
\verb"> end do: end do: end do:"\\}
On the next step we can construct the  operator  $N^*$ and its
minimal polynomial
 $\Delta(\lambda)$\\
{\footnotesize
\verb"> N:=evalm(inverse(P1)&*P2);"\\
\verb"> ed2:=array(identity,1.. 2*n,1.. 2*n):"\\
\verb"> Delta:=collect(simplify(sqrt(factor("\\
\verb">        det(N-lambda*ed2))),symbolic),lambda,factor);"\\
\verb"> if coeff(Delta,lambda,n)<0 then Delta:=-Delta: end if:"\\}
The coefficients of the minimal polynomial $\Delta(\lambda)$ are
entries of the control matrix  $F$
(\ref{f1-pedr})\\
{\footnotesize
\verb"> sigma:=vector(n):"\\
\verb"> for k to n do sigma[k]:=-coeff(Delta,lambda,n-k); end do:"\\
\verb"> F:=delcols(augment(sigma,ed),n+1.. n+1):"\\}
Now we introduce natural Hamilton function $H=H_1$ and unknown
integrals of motion $H_2,\ldots,H_n$ and calculate the
corresponding differentials $dH_k$\\
{\footnotesize
\verb"> H1:=add(p||k^2,k=1.. n)+V(q);"\\
\verb"> for k from 2 to n do H||k:=h||k(var); dH||k:=vector(2*n): end do:"\\
\verb"> for k to n do"\\
\verb">  for i to 2*n do dH||k[i]:=simplify(diff(H||k,var[i]),symbolic):"\\
\verb"> end do: end do:"\\}
On the final step after building  of the equations (\ref{F-pedr})\\
{\footnotesize
\verb"> ListEqH:=NULL:"\\
\verb"> for i to n do"\\
\verb">   Z||i:=simplify(evalm(N&*dH||i-add(F[i,j]*dH||j,j=1.. n))):"\\
\verb">   ListEqH:=ListEqH,seq(Z||i[k],k=1.. 2*n):"\\
\verb"> end do:"\\}
we solve the complete overdetermined system of algebro-differential equations\\
{\footnotesize
\verb"> Ans:=pdsolve({ListEq,ListEqH},{f1(q),V(q),seq(H||k,k=2.. n)},"\\
\verb">                               parameters={seq(c||j,j=1..8)});"\\}

\par\noindent
This programm allows us to construct three integrable systems listed
in the Proposition \ref{prop-lin} in a few second on the standard
personal computer. A little large programm allows us to construct
all the systems listed in this paper in  a few minutes.

\begin{rem} In order to cut the computer time and to exclude the trivial solutions
we have to add some simple equations to our system. For instance we
added the following inequalities  $\frac{\partial}{\partial p_i}
H_k\neq 0 $ and $\frac{\partial^2}{\partial q_1\partial
q_2}V(q_1,q_2)\neq 0$.
\end{rem}

\section{Conclusion}
\setcounter{equation}{0}

Using some recent results on the separation of variables for
bi-hamiltonian manifolds we propose a new method for construction on
the natural integrable systems. In this method the desired integrals
of motion are solutions of the overdetermined system of
algebro-differential equations, which arise from the Jacobi
identities for the Poisson pencil (\ref{jac1},\ref{jac2}) and from
invariance of the corresponding bi-Lagrangian  distribution with
respect to recurrence operator $N$ (\ref{F-pedr}). The algorithm may
be easy realized in any modern symbolic computational system. One of
the possible implementations may be found in the Appendix. So the
search for natural integrable systems is reduced to simple computer
calculations.

The main advantage of the proposed method is that we construct
integrals of motion of a natural integrable system, the
corresponding Poisson pencil and a part of the separated variables
simultaneously.

The main disadvantage  is that we have to use undefined polynomial
anzats for the second Poisson tensor $P^{\,\prime}$ and to choose
some control matrix $F$. So, we construct special families  of the
integrable systems associated with the a given $P^{\,\prime}$ and
$F$.

Of course, by using very simple substitutions
(\ref{lin-gen},\ref{gen-2})  we reproduce a majority of the known
natural integrable systems on the plane with cubic and quartic
integrals of motion. Nevertheless we did not get all the known
integrable systems \cite{hiet87}. Moreover we do not know how to
construct similar polynomials $P^{\,\prime}$ on the sphere or on the
ellipsoid.

The second  disadvantage is an exponential growth of the computer
time with the growth of  $n=\mbox{rank}\,P^{\,\prime}$ and with the
growth of the higher power of the matrix polynomial $P'$.

The author want to thank I.V. Komarov and M. Pedroni for valuable
discussions.

\end{document}